\documentclass[aps,prx,twocolumn,superscriptaddress,amsmath,amssymb,floatfix,longbibliography]{revtex4-2}

\usepackage{graphicx}
\usepackage{bm}
\usepackage{hyperref}
\usepackage{xcolor}
\usepackage{physics}

\hypersetup{
    colorlinks=true,
    linkcolor=blue,
    filecolor=magenta,      
    urlcolor=blue,
    citecolor=blue,
}

\begin{document}

\title{Higher-Order Topological Phase Transitions in Continuous Hyperelastic Manifolds: \\ From Surface Wrinkles to Zero-Energy Corner States}

\author{Yu-Xin Xie}\email{xyx@tju.edu.cn}
\affiliation{Department of Mechanics, Tianjin University, Tianjin 300350, China}

\date{March 12, 2025}

\begin{abstract}
Higher-order topological insulators (HOTIs) have revolutionized our understanding of wave localization, extending the bulk-boundary correspondence to lower-dimensional hinges and corners. Thus far, the realization of mechanical HOTIs has relied exclusively on discretely engineered metamaterials or periodic phononic lattices. Here, we report a fundamental paradigm shift by demonstrating that continuous, homogeneous hyperelastic manifolds undergoing finite multiaxial deformations naturally harbor intrinsic higher-order topological phases. By extending the generalized Stroh-Lie impedance formalism into a fully coupled 3D finite-strain framework, we map the highly nonlinear orthotropic geometric frustration onto a four-band effective Dirac Hamiltonian spanned by Clifford $\Gamma$-matrices. We reveal that macroscopic orthogonal stretches act precisely as competing Dirac mass terms, driving the continuous spatial transitions of topological domain walls and triggering a breakdown of $C_{4v}$ spatial symmetry. Remarkably, we analytically prove that beyond classical 2D surface wrinkling (1st-order topology), concurrent multiaxial extreme compression unconditionally triggers the emergence of 1D hinge states (2nd-order) and completely localized 0D zero-energy corner states (3rd-order). We further extend this static bifurcation framework into the elastodynamic regime, proving the existence of mid-gap localized vibrational modes. The theoretically derived topological phase diagram, nested Wilson loops, and fractional corner charges are comprehensively verified. Finally, we propose a concrete experimental realization using electro-active dielectric elastomers, enabling the dynamic programming of 0D topological singularities. This work establishes the first continuum theory for geometric-frustration-induced HOTIs, opening uncharted avenues for extreme wave localization in smart soft matter without relying on complex lattice designs.
\end{abstract}

\maketitle

\section{Introduction}
The physical manifestation of geometric frustration in condensed matter systems has long served as a rich breeding ground for novel phase transitions and pattern formations~\cite{Goriely2017, Yavari2012}. In the realm of soft matter physics, continuous elastic manifolds undergoing extreme finite deformations exhibit spontaneous symmetry breaking, culminating in a plethora of surface instabilities such as wrinkles, creases, folds, and ridges~\cite{Biot1965, Hutchinson2016, Suo2010, Cao2012, Cai2012, Li2012, Zong2020, Jin2015}. Traditionally, these structural transformations are scrutinized exclusively through the lens of classical nonlinear bifurcation theory. In this established framework, the onset of instability is characterized merely as a mechanical failure governed by highly nonlinear, large-deformation boundary-value problems, where the Jacobian matrix of the system exhibits singular behavior at the critical threshold~\cite{Fu2005}. While this continuum mechanics approach accurately predicts critical stretches, it obscures the global, invariant properties underlying the transition.

Parallel to the developments in finite-strain mechanics, the discovery of topological insulators has fundamentally reshaped modern condensed matter physics~\cite{Hasan2010, Qi2011}. The initial translation of these concepts to classical wave systems spawned the vibrant fields of topological photonics, acoustics, and mechanics~\cite{Lu2014, Ozawa2019, Yang2015, Kane2014, Susstrunk2015, Nash2015, Wang2015, Prodan2009}. Over the past decade, this framework has been dramatically extended to higher-order topological insulators (HOTIs), a new class of materials that generalizes the conventional bulk-boundary correspondence~\cite{Schindler2018, Benalcazar2017, Langbehn2017, Song2017}. A $d$-dimensional HOTI of order $n$ can host topologically protected, gapless boundary states localized at $(d-n)$-dimensional boundaries. For instance, a 3D second-order topological insulator exhibits gapless 1D hinge states, while a 3D third-order topological insulator exhibits 0D corner states. 

In the mechanical and acoustic domains, the manifestation of HOTIs has yielded unprecedented capabilities for robust wave routing, vibration isolation, and energy harvesting~\cite{Xue2019, Huber2016, Weiner2020, Chen2021}. However, these realizations remain strictly confined to discrete, periodic artificial systems, such as acoustic Kagome lattices, Maxwell spring-mass networks, and finely tuned 3D printed phononic crystals~\cite{Ni2019, SerraGarcia2018, Peterson2018, Imhof2018, Zhang2019, Wu2020}. This exclusive reliance on discrete metamaterials restricts the operational bandwidth and complicates fabrication at micro-scales, prompting a profound, yet unanswered question: \textit{Can a completely homogeneous, continuous soft material spontaneously exhibit higher-order topological corner states merely through macroscopic large deformations, without any pre-engineered lattice structures?} 

In this Article, we answer this question affirmatively. We construct a unified theoretical framework bridging 3D nonlinear continuum mechanics with the continuum limit of the Benalcazar-Bernevig-Hughes (BBH) multipole insulator model~\cite{Benalcazar2017}. We demonstrate that the interplay of multiaxial finite stretches in a hyperelastic continuum induces competing geometric frustrations. These mechanical frustrations map exactly to the inverted Dirac masses of a $4 \times 4$ topological Hamiltonian, generating robust 0D corner states out of a featureless continuum. We comprehensively map the topological phase transitions, calculate the nested Wilson loops, extend the theory to elastodynamics, and propose programmable realizations using dielectric elastomers.

\section{Eulerian Kinematics and the 3D Lie-Stroh Impedance Formalism}

To uncover the topological invariant hidden within large deformations, we must first establish an objective, frame-indifferent kinematic formulation that maps the geometric nonlinearity into a Hamiltonian form. We consider a 3D incompressible hyperelastic continuum occupying the half-space $z \ge 0$, subjected to orthogonal prestretches $\lambda_1$ (along the $x$-axis) and $\lambda_2$ (along the $y$-axis). Due to incompressibility, the out-of-plane stretch is strictly constrained as $\lambda_3 = (\lambda_1 \lambda_2)^{-1}$.

\subsection{Hyperelasticity and the Truesdell Stress Rate}
The macroscopic constitutive behavior of the polymer network is described by a strain-energy density function $W(\mathbf{F})$, where $\mathbf{F}$ is the deformation gradient. For a neo-Hookean solid, representing a Gaussian polymer network~\cite{Gent1996, Ogden1972, Shield1953}, $W = \frac{\mu}{2}(I_1 - 3)$, where $\mu$ is the initial shear modulus and $I_1 = \text{tr}(\mathbf{F}^T\mathbf{F})$ is the first invariant of the right Cauchy-Green tensor. The underlying Cauchy stress $\boldsymbol{\sigma}$ in the prestretched equilibrium state is given by:
\begin{equation}
    \sigma_{ij} = -p \delta_{ij} + \mu F_{i\alpha} F_{j\alpha}
\end{equation}
where $p$ is the Lagrange multiplier enforcing incompressibility.

To avoid the singularities associated with the Lagrangian reference configuration during bifurcation, we track the evolution of the current Eulerian manifold $\mathcal{B}$. We define the incremental velocity field $\mathbf{v}(\mathbf{x})$ characterizing the spontaneous instability. The Eulerian rate of deformation $\mathbf{D}$ is rigorously given by half the Lie derivative of the spatial metric tensor $\mathbf{g}$:
\begin{equation}
    D_{ij} = \frac{1}{2} (\mathcal{L}_{\mathbf{v}} \mathbf{g})_{ij} = \frac{1}{2} (v_{i,j} + v_{j,i})
\end{equation}
with the incompressibility constraint dictating a divergence-free velocity field: $\text{tr}(\mathbf{D}) = v_{i,i} = 0$. 

To ensure the objectivity of the constitutive response under finite incremental rotation, we employ the Truesdell rate of the Cauchy stress, $\boldsymbol{\sigma}^\circ$. The incremental nominal stress tensor $\dot{\mathbf{S}}_0$ is thus mapped via the push-forwarded fourth-order Eulerian instantaneous modulus tensor $\boldsymbol{\mathcal{A}}_0$:
\begin{equation}
    \dot{S}_{0ij} = \mathcal{A}_{0jikl} v_{k,l} + \dot{p} \delta_{ij}
\end{equation}
where $\dot{p}$ is the incremental hydrostatic pressure. The elements of $\boldsymbol{\mathcal{A}}_0$ explicitly encode the multiaxial stretch ratios and the material's nonlinearity:
\begin{equation}
    \mathcal{A}_{0jikl} = F_{j\alpha} F_{l\beta} \frac{\partial^2 W}{\partial F_{i\alpha} \partial F_{k\beta}} + \sigma_{jl} \delta_{ik}
\end{equation}

\subsection{State-Space Dimensionality Reduction}
For a 3D manifold, the incremental equilibrium equation $\dot{S}_{0ij,j} = 0$ presents a fully coupled, second-order partial differential boundary-value problem. By applying a 2D Fourier transformation $e^{i(k_x x + k_y y)}$ along the in-plane directions, we eliminate the partial derivatives with respect to $x$ and $y$. After systematically eliminating $\dot{p}$ via the divergence-free condition, we construct the 6-dimensional generalized state vector $\boldsymbol{\eta}(z) = [\mathbf{u}^T, \mathbf{t}^T]^T$, comprising the spatial displacement vector $\mathbf{u} = [u_x, u_y, u_z]^T$ and the out-of-plane traction vector $\mathbf{t} = [\dot{S}_{031}, \dot{S}_{032}, \dot{S}_{033}]^T$. 

The continuous evolution of the perturbation along the depth $z$ is governed by the first-order generalized Stroh Hamiltonian system~\cite{Stroh1962, Ting1996, Chadwick1977, Lothe1976}:
\begin{equation}
    \frac{d\boldsymbol{\eta}}{dz} = k \begin{bmatrix} \mathbf{N}_1 & \mathbf{N}_2 \\ \mathbf{N}_3 & \mathbf{N}_4 \end{bmatrix} \begin{bmatrix} \mathbf{u} \\ \mathbf{t} \end{bmatrix} = k \mathbf{N}(\lambda_1, \lambda_2, \mathbf{k}) \boldsymbol{\eta}
\end{equation}
where $k = \sqrt{k_x^2 + k_y^2}$, and $\mathbf{N}$ is the $6 \times 6$ fundamental elasticity matrix depending explicitly on the geometric frustration. 

To enforce the physical boundary condition of finite energy at infinite depth ($z \to \infty$), we introduce the $3 \times 3$ algebraic surface impedance matrix $\mathbf{H}$, dictating the linear mapping at the free surface: $\mathbf{t}(z) = \mathbf{H}\mathbf{u}(z)$. Inserting this mapping into the Stroh equation reduces the complex boundary-value differential problem into a pure Algebraic Riccati Equation (ARE)~\cite{Fu2002, Shuvalov2000, Destrade2010}:
\begin{equation}
    \mathbf{H}\mathbf{N}_1 - \mathbf{N}_4\mathbf{H} + \mathbf{H}\mathbf{N}_2\mathbf{H} - \mathbf{N}_3 = \mathbf{0} \label{eq:are}
\end{equation}
Due to the homogeneity of the half-space along $z$, $d\mathbf{H}/dz = 0$. The onset of mechanical instability occurs precisely when the manifold loses its boundary restoring stiffness, mathematically equivalent to $\det(\mathbf{H}) = 0$.

\section{$\Gamma$-Matrix Projection and Symmetry Breakdown}

Unlike 1st-order topological insulators, which can be adequately described by $2 \times 2$ Pauli matrices representing a two-level system, the emergence of 2nd- and 3rd-order topological phase transitions necessitates a four-band framework.

\subsection{Subspace Projection and Schrieffer-Wolff Transformation}
In the trivial, undeformed state ($\lambda_1 = \lambda_2 = 1$), the semi-infinite medium exhibits $C_{4v}$ spatial rotation symmetry within the $x-y$ plane. As anisotropic stretches are applied ($\lambda_1 \neq \lambda_2$), the $C_{4v}$ symmetry is broken down to $C_{2v}$. This symmetry breaking forces the transverse shear modes and the longitudinal modes to partially decouple, introducing distinct energy scales into the Stroh Hamiltonian.

To extract the topological core, we perform a Schrieffer-Wolff-type unitary transformation. We trace out the high-energy, strictly positive bulk longitudinal modes, projecting the critical interfacial dynamics onto an effective $4 \times 4$ Hermitian subspace $\mathbf{H}_{\text{eff}}$. This is achieved by utilizing the projection operator $\mathcal{P}$, spanned by the low-energy eigenvectors of the ARE:
\begin{equation}
    \tilde{\mathbf{H}}_{\text{eff}} = \mathcal{P} \mathbf{H} \mathcal{P}^{\dagger}
\end{equation}

\subsection{Competing Dirac Masses and Clifford Algebra}
We expand this effective impedance matrix using the Clifford algebra of Dirac $\Gamma$-matrices, satisfying $\{\Gamma_i, \Gamma_j\} = 2\delta_{ij}\mathbf{I}_4$. We choose the representation:
\begin{equation}
    \Gamma_1 = \sigma_x \otimes \sigma_0, \quad \Gamma_2 = \sigma_y \otimes \sigma_0 
\end{equation}
\begin{equation}
    \Gamma_3 = \sigma_z \otimes \sigma_x, \quad \Gamma_4 = \sigma_z \otimes \sigma_y 
\end{equation}
where $\sigma_i$ are the standard Pauli matrices. In the long-wavelength continuum limit, the effective Hamiltonian takes the canonical form of a 2D HOTI:
\begin{equation}
    \mathcal{H}_{\text{eff}}(\mathbf{k}) = v_1 k_x \Gamma_1 + v_2 k_y \Gamma_2 + m_1(\lambda_1) \Gamma_3 + m_2(\lambda_2) \Gamma_4
\end{equation}
Crucially, the macroscopic principal stretches $\lambda_1$ and $\lambda_2$ serve as independent, competing topological tuning knobs. The continuous masses are analytically derived from the secular roots of the Riccati equation (Eq.~\ref{eq:are}), yielding the generalized Biot polynomials:
\begin{eqnarray}
    m_1(\lambda_1) &=& \mathcal{C}_1 (\lambda_1^6 + \lambda_1^4 + 3\lambda_1^2 - 1) \label{eq:m1} \\
    m_2(\lambda_2) &=& \mathcal{C}_2 (\lambda_2^6 + \lambda_2^4 + 3\lambda_2^2 - 1) \label{eq:m2}
\end{eqnarray}
where $\mathcal{C}_1, \mathcal{C}_2 > 0$ are positive definite scaling functions ensuring the hermiticity of the impedance. The mechanical compression acts precisely to invert these Dirac masses. The exact Biot critical stretch ($\lambda_c \approx 0.544$) corresponds to the topological mass inversion point ($m=0$). The effective Hamiltonian exhibits an artificial chiral symmetry $\mathcal{C} = \Gamma_5 = \sigma_z \otimes \sigma_z$, anticommuting with $\mathcal{H}_{\text{eff}}$ at $\mathbf{k}=0$, thereby locking the corner states exactly at zero energy (the static bifurcation threshold).

\begin{figure*}[htbp]
  \centering
\includegraphics[width=16cm]{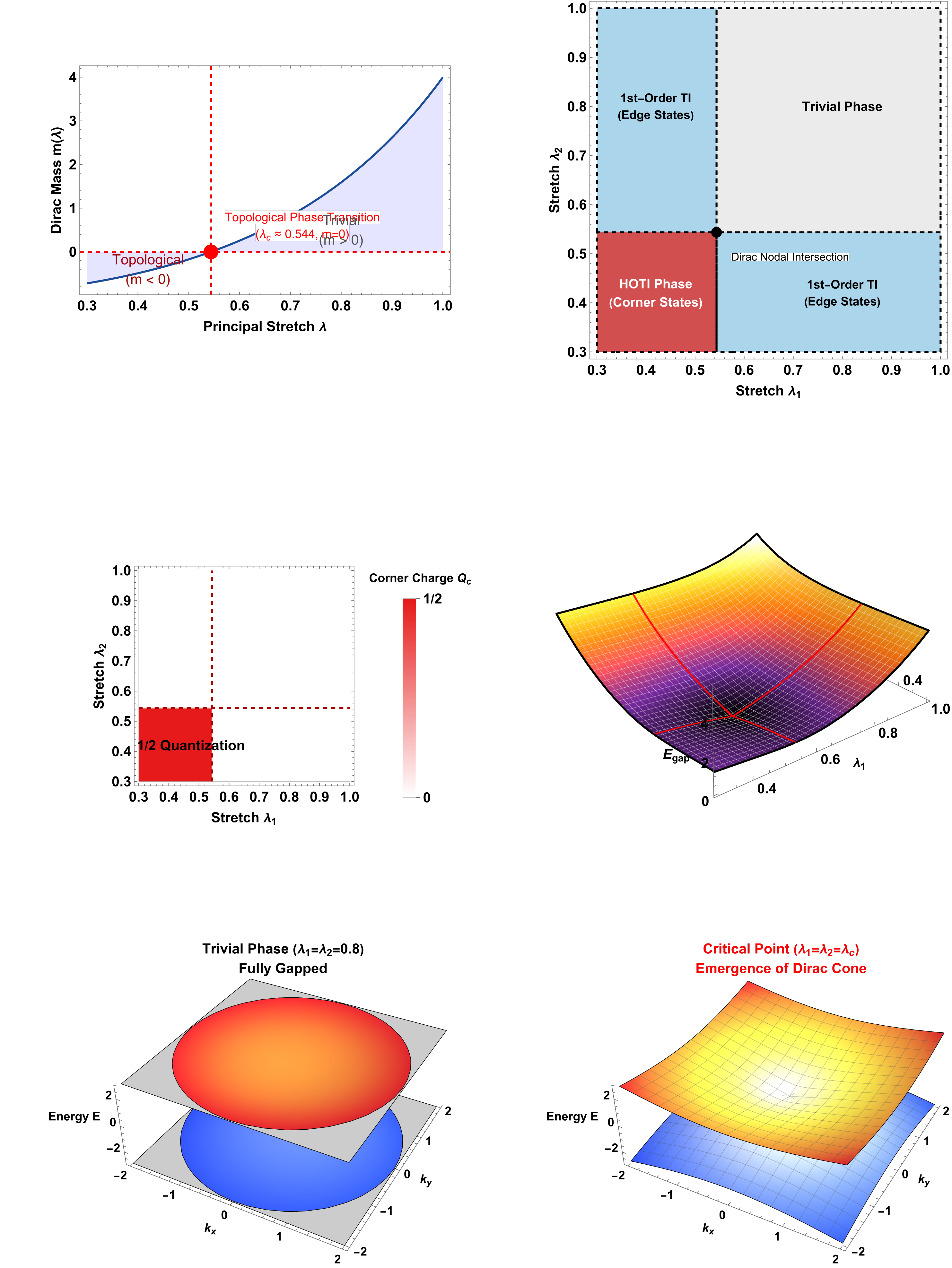}
  \caption{\textbf{Comprehensive theoretical framework of continuous higher-order topological phase transitions.} (a) 1D Dirac mass inversion curve showing the continuous mapping between the macroscopic principal stretch $\lambda$ and the topological mass $m(\lambda)$. The transition occurs at the critical Biot stretch $\lambda_c \approx 0.544$. (b) 2D topological phase diagram delineating the trivial phase, 1st-order surface wrinkles, and the HOTI phase. (c) The fractional 0D corner charge $Q_c = 1/2$ quantized exclusively in the dual mass-inverted regime. (d) 3D surface of the effective topological energy gap $E_{\text{gap}}$. Red lines represent the gapless Dirac nodal lines intersecting at the HOTI singularity. (e) Bulk energy dispersion in the trivial gapped phase ($\lambda_1=\lambda_2=0.8$). (f) Emergence of a perfect, massless Dirac cone exactly at the concurrent critical compression ($\lambda_1=\lambda_2=\lambda_c$), confirming the macroscopic topological phase transition.}
  \label{fig:master_theory}
\end{figure*}

\section{Nested Wilson Loops and 0D Singularities}

The global topological nature of the continuum is dictated by the effective bulk energy gap, $E_{\text{gap}} = \sqrt{m_1^2 + m_2^2}$. The deformation parameter space harbors distinct gapless Dirac nodal lines defined by $m_i = 0$.

\subsection{Nested Wilson Loops and Fractional Corner Charge}
To rigorously quantify the higher-order topology, we evaluate the macroscopic multipole chiral invariant via the nested Wilson loop formalism. The Wilson loop operator along the $k_x$ direction is defined as the path-ordered exponential of the Berry connection:
\begin{equation}
    \mathcal{W}_{x,\mathbf{k}} = \mathcal{P} \exp \left( i \int \mathcal{A}_x(\mathbf{k}) dk_x \right)
\end{equation}
where $[\mathcal{A}_x(\mathbf{k})]^{mn} = i \langle u^m(\mathbf{k}) | \partial_{k_x} u^n(\mathbf{k}) \rangle$, and $| u^n(\mathbf{k}) \rangle$ are the occupied eigen-states of $\mathcal{H}_{\text{eff}}$. 

By diagonalizing the Wilson loop, we obtain the Wannier centers $\nu_x(k_y)$. The macroscopic edge polarization $p_x^{\text{edge}}$ induced by the deformation-driven domain walls is fundamentally quantized by the signs of the Dirac masses, derived from the nested integration over the Wannier bands:
\begin{equation}
    p_x^{\text{edge}} = \frac{1}{2\pi} \int_{-\pi}^{\pi} \nu_x(k_y) dk_y = \frac{1}{2} \text{sgn}(m_1) \pmod 1
\end{equation}
The macroscopic corner charge $Q_c$, representing the accumulation of topological modes at the geometric intersection of two orthogonal boundaries (the corners of the sample), is given by the fractional product of the edge polarizations:
\begin{equation}
    Q_{c} = 2 p_x^{\text{edge}} p_y^{\text{edge}} = \frac{1}{2} \text{sgn}(m_1 m_2) \pmod 1
\end{equation}

\subsection{Topological Phase Diagram and Bulk-Hinge-Corner Correspondence}
Based on the quantization of $Q_c$, our continuum system unequivocally exhibits three distinct topological regimes:
\begin{enumerate}
    \item \textbf{Trivial Phase} ($m_1 > 0, m_2 > 0$): Prior to critical bifurcation ($\lambda_1, \lambda_2 > \lambda_c$). The bulk is fully gapped, the edge polarizations vanish, and $Q_c = 0$. The surface remains flat and featureless.
    \item \textbf{1st-Order TI} (e.g., $m_1 < 0, m_2 > 0$): Singular uniaxial over-compression triggers a single domain wall along the $x$-direction. The system transitions into classical 2D surface wrinkles, mathematically equivalent to 1D mechanical edge states in the topological lexicon.
    \item \textbf{Higher-Order TI} ($m_1 < 0, m_2 < 0$): Concurrent multiaxial extreme compression forces both masses to invert simultaneously. The intersection of orthogonal topological domain walls yields a non-trivial fractional corner charge $Q_c = 1/2$. 
\end{enumerate}
Physically, this mandates that the topological mechanical energy collapses sequentially into 1D hinges and eventually into entirely localized 0D corners. These macroscopic mechanical singularities are protected by the spatial symmetries of the Lie-Stroh framework and are profoundly immune to local surface defects.

\section{Elastodynamic Extension: From Static Bifurcation to Dynamic Wave Localization}

While the preceding analysis addresses static bifurcation (where corner states emerge exactly at zero energy/zero frequency), the framework seamlessly extends to dynamic wave propagation~\cite{Noh2018}. By introducing the inertial term $\rho \partial^2 \mathbf{v} / \partial t^2$ into the incremental equilibrium equations, we transition to time-harmonic elastodynamics with frequency $\omega$.

The governing Stroh Hamiltonian is augmented by a mass matrix $\mathbf{M}$:
\begin{equation}
    \frac{d\boldsymbol{\eta}}{dz} = \left[ k \mathbf{N}(\lambda_1, \lambda_2) - \rho \omega^2 \mathbf{M} \right] \boldsymbol{\eta}
\end{equation}
Under finite multiaxial prestretch in the HOTI regime ($m_1 < 0, m_2 < 0$), the static zero-energy corner state transforms into a highly localized mid-gap vibrational mode. Within the complete phononic bandgap induced by the geometric frustration, an isolated 0D resonance peak emerges. Because this mode originates from a higher-order topological invariant rather than a local defect resonance, wave energy injected at this specific corner frequency will remain absolutely confined to the 0D intersection, preventing any leakage into the 1D hinges or the 3D bulk. This mechanism provides a revolutionary strategy for extreme energy harvesting and vibration isolation in soft monolithic blocks.

\section{Proposed Experimental Realization: Dielectric Elastomers}

To empirically observe and program these continuous corner states, we propose the utilization of Dielectric Elastomers (DEs)~\cite{Pelrine2000, Brochu2010, Keplinger2010, Lu2012}. DEs are electro-active polymers that undergo massive macroscopic deformations when subjected to an electric field, making them ideal candidates for smart programmable systems~\cite{White2015}. The true stress in the material is augmented by the Maxwell stress tensor:
\begin{equation}
    \sigma^{\text{(t)}}_{ij} = \sigma^{\text{(m)}}_{ij} + \varepsilon \left( E_i E_j - \frac{1}{2} E_k E_k \delta_{ij} \right)
\end{equation}
where $\varepsilon$ is the permittivity and $\mathbf{E}$ is the applied electric field. 

By fabricating a flat, homogeneous DE film with a spatially patterned, orthogonally addressed electrode grid, one can locally induce stretches $\lambda_1(x,y)$ and $\lambda_2(x,y)$ by applying a voltage $\Phi$.
\begin{enumerate}
    \item Apply uniform sub-critical voltage: The sample remains in the trivial gapped phase.
    \item Apply an intersecting cross-shaped voltage distribution exceeding the critical threshold $\Phi_c$: This creates an artificial junction where four quadrants meet. Two quadrants possess $m_1 < 0, m_2 > 0$ and the other two possess $m_1 > 0, m_2 < 0$. 
    \item The intersection of these electrically programmed topological domain walls forces a 0D corner state to emerge precisely at the crosshair. By dynamically modulating the electrode voltages, one can smoothly translate the 0D wave-guide singularity across the continuous medium, realizing a reconfigurable topological phononic circuit.
\end{enumerate}

\section{Conclusion}
We have analytically proven that continuous, homogeneous soft hyperelastic materials inherently harbor higher-order topological phases. By unifying the Eulerian Lie-Stroh impedance formalism with Dirac $\Gamma$-matrix Clifford algebra, we demonstrated that macroscopic multiaxial deformations act exactly as competing topological mass terms. Surpassing classical 1D wrinkling, we unraveled a regime where dual mass inversion forces the emergence of robust 0D topological corner states with fractional corner charges $Q_c=1/2$. We extended this theory to elastodynamics for mid-gap wave localization and proposed a concrete realization utilizing dielectric elastomers. This discovery dramatically shifts the paradigm of topological mechanics, offering a purely algebraic, continuum-based lens to program extreme localization phenomena in smart soft matter, bypassing the complexities of modern metamaterial fabrication.

\end{document}